\documentstyle [pre,aps,epsfig,multicol]{revtex}
\newenvironment{mytitle}{\begin{center} \large \bf }{\\ [.1in]\end{center}}
\newenvironment{myauthor}{\begin{center} \large }{\\ [.1in]\end{center}} 
\newenvironment{myinstit}{\begin{center} \large \it}{\end{center}}
\begin{document}

\thispagestyle{empty}

\begin{mytitle}
{\LARGE \bf Chaotic Waveguide-Based Resonators for Microlasers}
\end{mytitle}

\begin{myauthor}
J. A. M\'endez-Berm\'udez$^1$, G. A. Luna-Acosta$^1$, P. \v{S}eba$^{2,3}$, 
and K. N. Pichugin$^{2,4}$
\end{myauthor}

\begin{myinstit}
$^1$Instituto de F\'{\i}sica, Universidad Aut\'onoma de Puebla, Apartado Postal J-48, Puebla 72570, M\'exico\\
$^2$Department of Physics, University Hradec Kralove, Hradec Kralove, Czech Republic\\
$^3$Institute of Physics, Czech Academy of Sciences, Cukrovarnicka 10, 
Prague, Czech Republic\\
$^4$Kirensky Institute of Physics, 660036 Krasnoyarsk, Russia
\end{myinstit}

\begin{center}
{\bf Abstract} 
\end{center}
{\small
We propose the construction of highly directional emission microlasers using two-dimensional high-index semiconductor waveguides as {\it open} resonators. The prototype waveguide is formed by two collinear leads connected to a cavity of certain shape. The proposed lasing mechanism requires that the shape of the cavity yield mixed chaotic ray dynamics so as to have the appropiate (phase space) resonance islands. These islands allow, via Heisenberg's uncertainty principle, the appearance of quasi bound states (QBS) which, in turn, propitiate the lasing mechanism. The energy values of the QBS are found through the solution of the Helmholtz equation. We use classical ray dynamics to predict the direction and intensity of the lasing produced by such open resonators for typical values of the index of refraction.\\

}
\begin{multicols}{2}

Studies on microcavities which serve as optical resonators have led to the design of semiconductor microlasers, see for example \cite{8}. One way to obtain high performance microcavities is through the excitation of {\it Wispering-Gallery Modes} (WGM), first observed in 1961 using a spherical sample of $CaF_2:Sm^{2+}$ \cite{1}. WGM living in microcavities with spherical (or circular) geometry were studied in polyestyrene spheres \cite{1.1}, spherical doplets \cite{2}, spheres of fused quartz \cite{4}, and in semiconductor disks \cite{6}. In Ref. \cite{1} it was also suggested that WGM could be tuned by deforming the cavity from a spherical shape. This idea has been explored in spheroidal doplets \cite{7}, fused silica spheroids \cite{11}, deformed fused silica spheres \cite{20,36}, deformed disks \cite{38}, and other geometries \cite{19.26.27}. The WGM in a deformed disc were first analyzed in terms of chaos theory in Ref. \cite{10}, and later a similar analysis was made for deformed doplets \cite{14}. Furthermore, based on features of nonlinear dynamics, the {\it asymetric resonant cavity} (ARC) as a resonator for semiconductor lasers was proposed \cite{arc1}. The ARCs are two-dimensional (2D) circular resonators with quadrupolar deformations. Small deformations from circular geometry were shown to spoil the $Q$ of WGM but it was found that for higher deformations {\it bow-tie}-shaped resonances, instead of WGM, are responsible for the improvement in the power and directionality of the laser emmission \cite{arc2}.

Here we propose the use of a 2D waveguide system as an {\it open} resonator with similar topological properties as those of the ACRs with large deformations. We believe that the great advantage of an open resonator may be in that no pedestals or couplers (pumpers) close to the cavity are needed, as is the case with the currently investigated 3D and 2D microlasers, see e.g., Refs. \cite{11,6,38,arc2}.\

The waveguide system is formed by a cavity connected to two collinear semi-infinite leads of width $d$ extended along the $x$-axis. Our prototype waveguide has the geometry of the so-called {\it cosine billiard} extensively studied in \cite{infinite,finite1,finite2,ketz1,ketz2}: it has a flat wall at $y=0$ and a deformed wall given by $y(x)=d+a [1-\cos(2\pi x/L)]$, where $a$ is the amplitude of the deformation, and $L$ is the length of the cavity. 

We shall now discuss dynamical features of this system using, firstly, ray optics since it is assumed that the wavelength of light $\lambda$ is much smaller than the amplitude $a$ of the cosine profile. 

The set of parameters $(d,a,L) = (1.0,0.305,5.55)$ is known to produce mixed chaotic dynamics \cite{finite2,ketz1,ketz2} (an incomplete ternary horseshoe \cite{finite2}). The complete panorama of the ray dynamics is given by the Poincar\`{e} Map (PM), which is the plot of the orbits of a set of representative initial conditions as these cross, in a specified direction, a given surface of section in phase space \cite{note2}. For our purposes the surface of section can be either the botton (flat) or the top (cosine) boundary. Thus, each time a ray impinges on the top wall we plot the position $x$ and the sine of $\chi$, where $\chi$ is the angle the ray makes with the normal to the top boundary, see Fig. 1(a). Similarly we obtain another plot, see Fig. 1(b), when we use the bottom wall as a surface of section. Small variations in the geometrical parameters yield qualitatively the same topology in phase space. In Fig. 1(a) we clearly identify three islands: one large at $\sin(\chi)=0$ and two smaller ones centered at $\sin(\chi)=0.3$. Motion within these islands correspond to bounded motion inside the cavity ({\it i.e.} not accesible to scattering trajectories) and are produced by rays bouncing in the neighborhood of stable period-one and period-four fixed points, respectively \cite{note1}. With the surface of section at the top boundary, the phase space topology of this cavity is very similar to that of the ARC with large deformations \cite{arc2}. 
Observe that even though both figures, 1(a) and 1(b), are constructed using the same orbits in position space, their phase space topology are different. As shown below, we shall take advantage of this difference.

\vspace{-0.18in}
\begin{figure}[htb]
\begin{center}
\epsfig{file=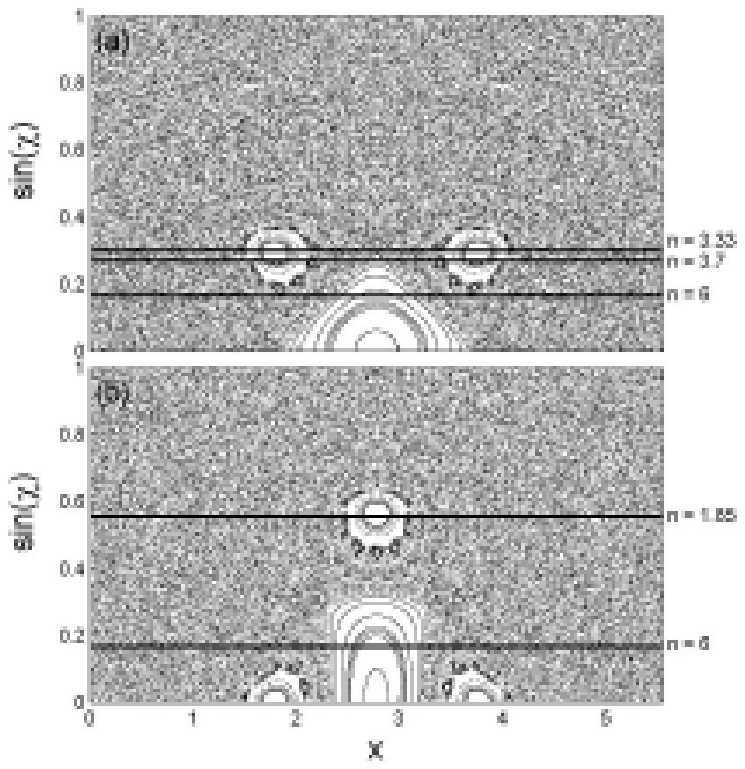,width=3.5in,height=3.5in}
\label{weaal0}
\end{center}
\end{figure}
\vspace{-0.46in}
{\small FIG. 1 Poincar\`{e} Maps for the closed cavity. The Poincar\`{e} surface of section is located at (a) the deformed wall: $y(x)=d+a [1-\cos(2\pi x/L)]$, and (b) the flat wall: $y(x)=0$. The parameters $(d,a,L) = (1.0,0.305,5.55)$ were used. Some critical lines for refractive scape are also shown.}\\

To study the wave scattering phenomena we numerically solve the Schr\"{o}dinger equation for our waveguide, as in \cite{finite1,finite2}, since we know \cite{stock,from} that the problem of a TM wave inside a 2D waveguide with Dirichlet boundary conditions (Helmholtz equation) is equivalent to that of a quantum wave in a 2D billiard (Schr\"{o}dinger equation).

It has been shown \cite{finite2,ketz1} for this waveguide with geometrical parameters $(d,a,L) = (1.0,0.305,5.55)$ that a plot of the Landauer conductance $G$ presents sharp resonances at energies where the wave functions either, tunnel into classical resonance islands (forbiden phase space regions for classical scattering), or have their support arround them. Those scattering wave functions that tunnel into classical islands are very similar to a set of eigenfunctions of the corresponding closed cavity \cite{ketz2}, hence we shall refer to them as quasi bound states (QBS). It was also found \cite{ketz2} that QBS may be of two types only: M-type or I-type, see Fig. 2. The M-type (I-type) corresponds to a wave function with support on the period-four (period-one) resonance. These patterns are produced by waves incident from the left, modes 3 and 6 respectively. Note that there is no evidence of reflection, which is a general feature of the lowest modes, as shown in Refs. \cite{finite1,finite2}.

In Fig. 3 we present a plot of $G$ for a range of energies supporting 20 propagating modes. The insets (a) and (b) enlarge two sharp dips of $G$ corresponding to the QBS of Fig. 2(a) and 2(b), respectively.

\begin{figure}[htb]
\begin{center}
\epsfig{file=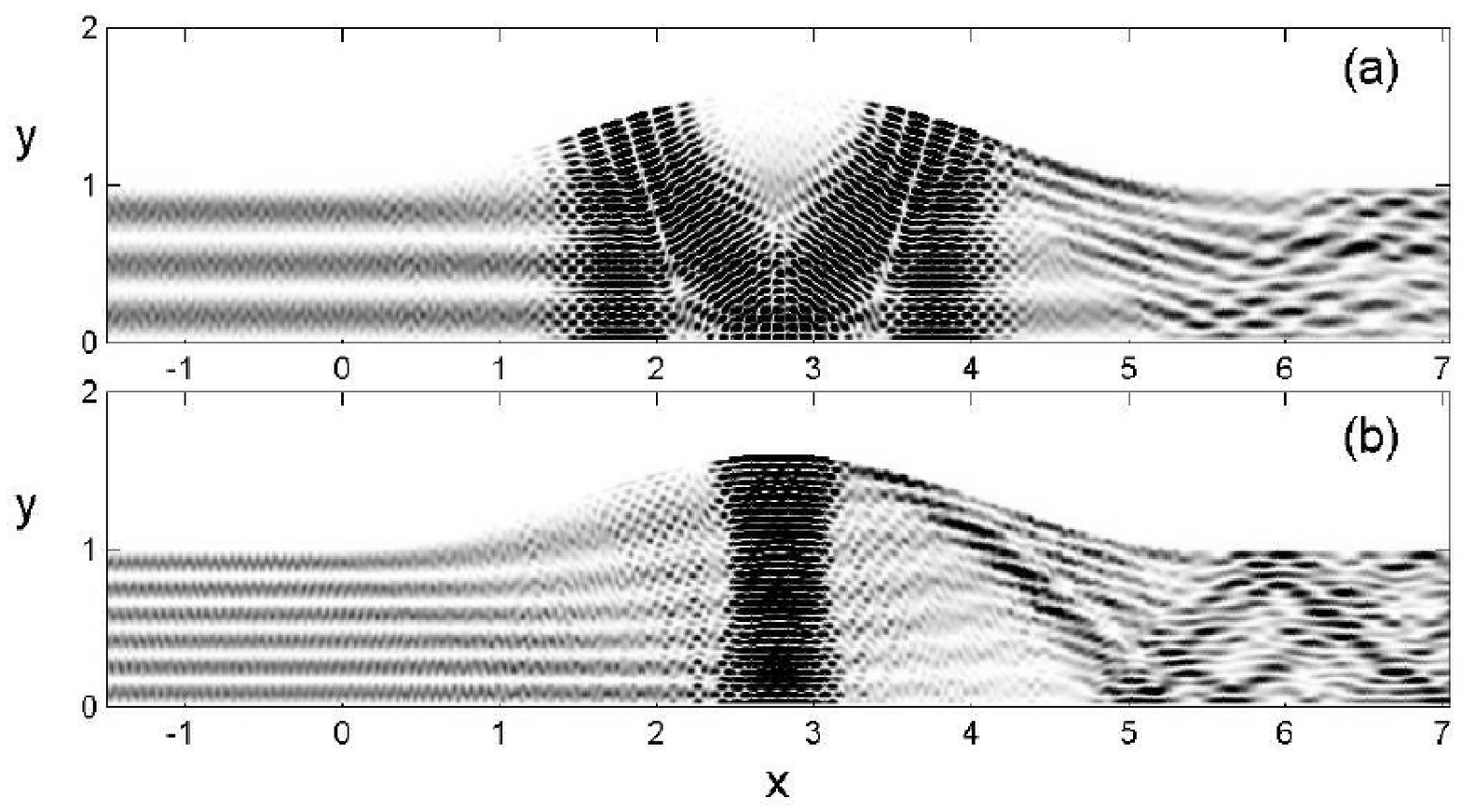,width=3.3in,height=2.1in}
\label{weaal0}
\end{center}
\end{figure}
\vspace{-0.28in}
{\small FIG. 2 Wave functions $\mid \Psi(x,y) \mid^2$ calculated at resonance. (a) $E=4008.8583$, (b) $E=4027.0597$. These QBS correspond to modes (a) 3, and (b) 6. The cavity is defined in the range $x=[0,5.55]$. $E=E_L/E^* \sim M_{max}^2\pi^2$; where $E_L$ is the expression for the energy in the leads, $E^*=\hbar^2/(2m_ed^2)$, and $M_{max}$ is the largest transversal mode $m$ beyond which the longitudinal wave vector 
$k_m =\sqrt{(2m_eE_L/\hbar^2)-(m^2 \pi^2/d^2)}$ becomes complex.}

\begin{figure}[htb]
\begin{center}
\epsfig{file=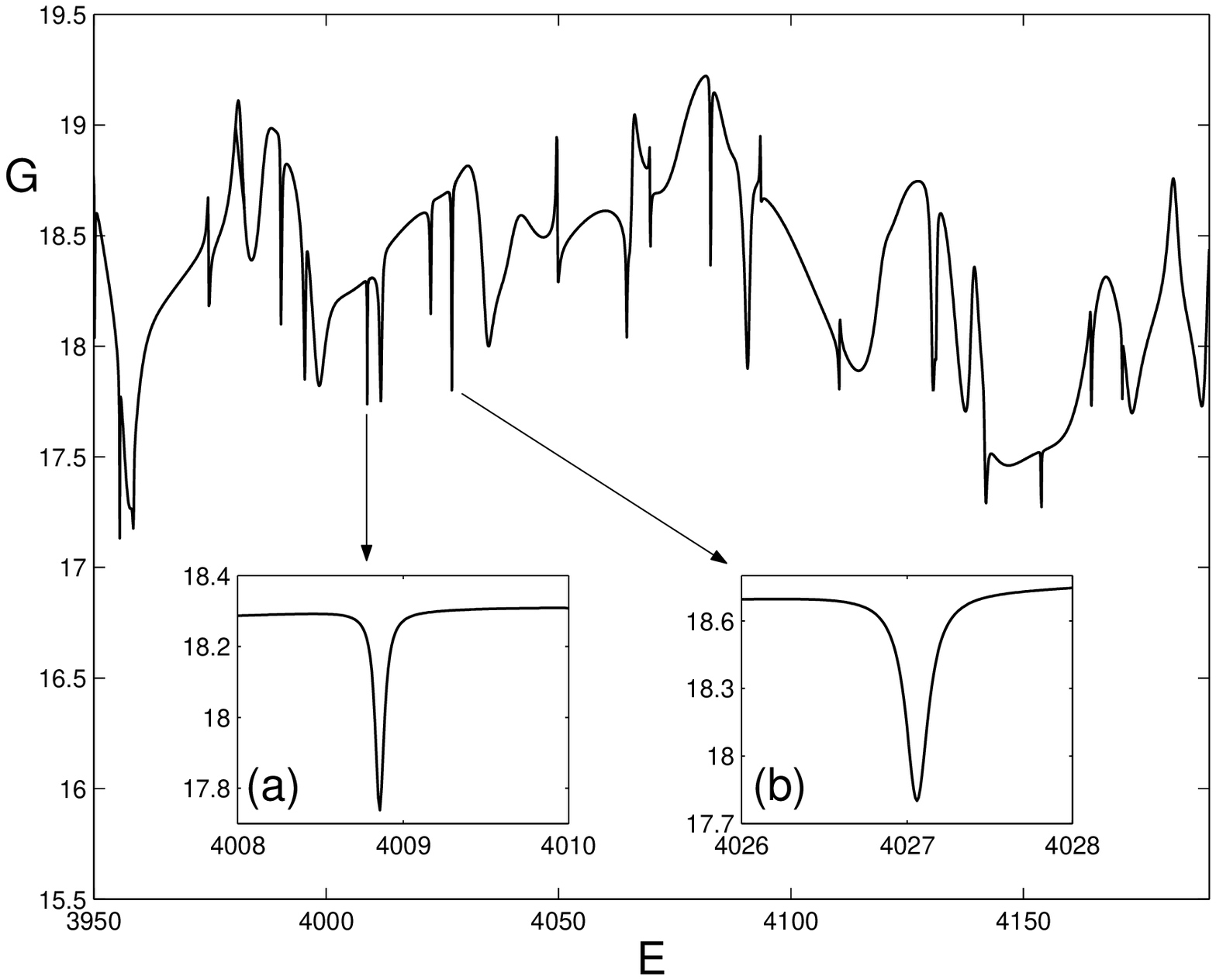,width=3.3in,height=2.5in}
\label{weaal0}
\end{center}
\end{figure}
\vspace{-0.28in}
{\small FIG. 3 Dimensionless conductance $G$ as a function of $E$. Insets (a) and (b) enlarge the dips whose energies where used to calculate the wave functions of Figs. 2(a) and 2(b), respectively.}\\

If the waveguide system is constructed with a semiconductor material having a refraction index $n$ and we introduce a wave into one of the leads with an energy corresponding to a sharp resonance in $G$, the cavity acts as an {\it open} resonator since the light will bounce in the cavity region following an M-type or an I-type orbit. Also, we assume silvered surfaces along the leads so that light inside the waveguide undergoes especular reflections at the silvered surfaces and have a chance to escape only in the cavity region. According to Snell's law there is a critical angle, $\chi_c = \sin^{-1}(1/n)$ such that, if $\chi>\chi_c$ the ray light reflects specularly, while if $\chi<\chi_c$ the ray light refracts. In Fig. 1 some critical lines for refractive escape $\sin(\chi_c)$ are ploted for different refractive indices. For a given value of $n$, orbits corresponding to points above (below) the critical line will reflect (escape).

Since the QBS form M-type or I-type patterns we can use ray dynamics to estimate the direction (using Snell's law) and the intensity (using Fresnel transmitance and refractance) of the emission for a given refractive index $n$. We do this by following the motion of an ensamble of rays with initial conditions inside the resonance islands and by counting the number of times the rays visit each of the elements of a mesh defined both, inside and outside (to consider escape from) the cavity. The mesh is a $555 \times 555$ square lattice which spans the whole plots of Figure 4. At the end of the counting each mesh element with coordinates $(x,y)$ has a weight $\rho(x,y)$ which is the classical counterpart of the wave function $\mid \Psi(x,y) \mid^2$ \cite{lass}. In Fig. 4 we show the patterns obtained from ray dynamics for both M-type [Fig. 4(a-b)] and I-type [Fig. 4(c-d)] functions for different values of $n$.

\begin{figure}[htb]
\begin{center}
\epsfig{file=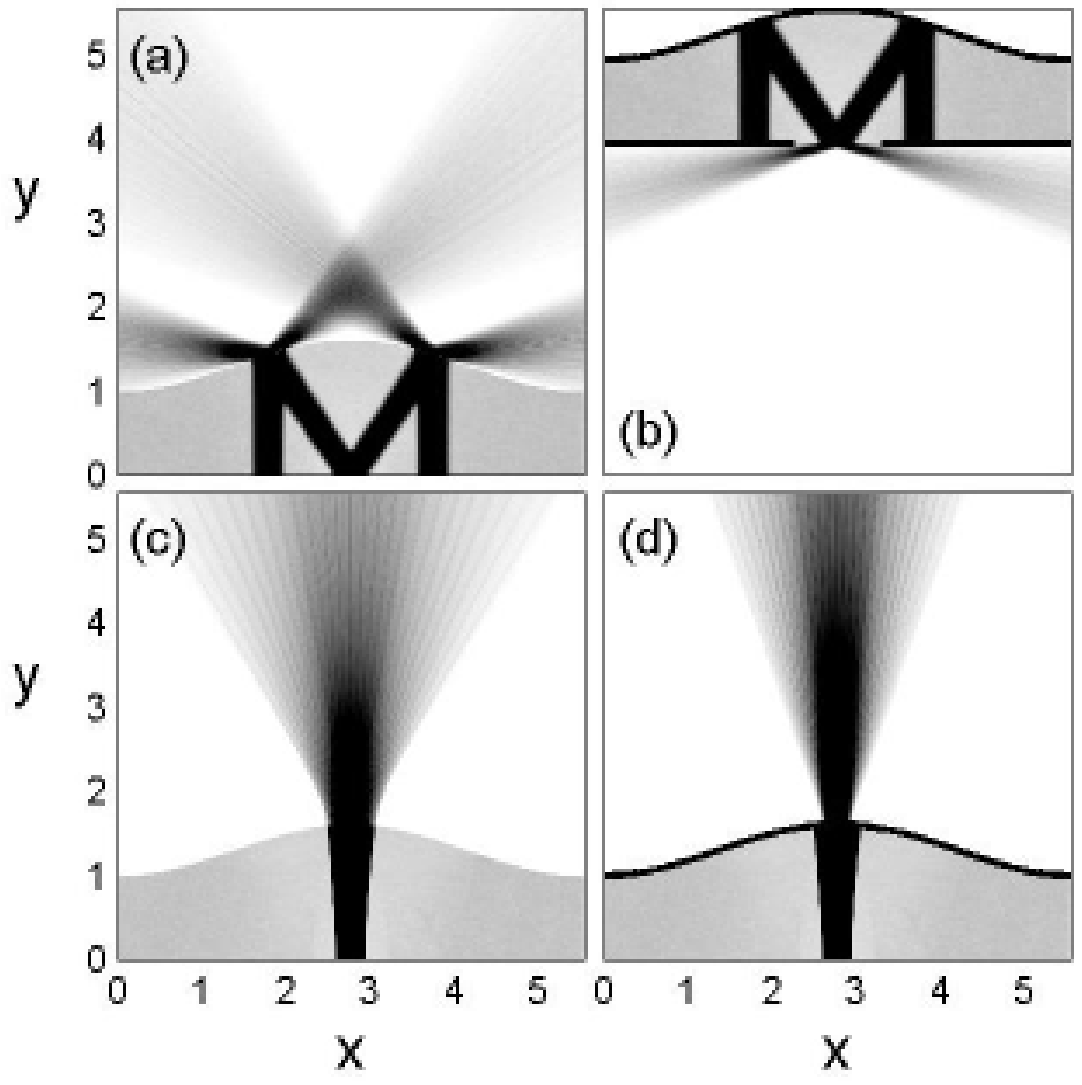,width=3.3in,height=3.3in}
\label{weaal0}
\end{center}
\end{figure}
\vspace{-0.28in}
{\small FIG. 4 Ray-optics prediction $\rho(x,y)$ of lasing emission for the M-type (a-b) and I-type (c-d) wave functions, where (a) $n=3.33$, (b) $n=1.85$, (c) $n=6$, and (d) $n=3.7$. The thick black lines in (b-d) means silvered surfaces.}\\

Consider the M-type QBS [Fig. 2(a)] that lives on the small islands of the PMs of Fig. 1, as shown in \cite{finite2}. Let us also assume that the flat surface of the cavity is silvered, so light bouncing inside the cavity may only escape through its deformed boundary. In this case the M-type QBS is equivalent to the bow-tie mode present in the ARCs with large deformations \cite{arc2}. As an illustration, if the waveguide is constructed with a semiconductor with refraction index $n=3.33$, the line for refractive escape, shown in Fig. 1(a), will cut the small islands centered at $\sin(\chi)=0.3$ approximately through the middle. Since the average angle of incidence of the M-type function on the deformed boundary is slightly less than the critical angle $\chi_c\approx 17.5^\circ$, the reflectivity $R$ of this boundary is (given by the Fresnel's coefficients) sufficiently high ($\sim 70$ percent) to permit laser action with the direction shown in Fig. 4(a). In contrast, if light is allowed to escape only through the flat boundary then the M-type function has two different angles of incidence on that boundary, see Fig. 1(b): one at $\chi_c\approx 90^\circ$ (the two small half-islands at $\sin(\chi)\approx 0$) and other at $\chi_c\approx 33^\circ$ (the small island centered at $\sin(\chi)\approx 0.57$). For a semiconductor cavity with $n<6$, the boundary presents a low $R$ ($<50$ percent) at normal incidence ($\chi_c\approx 90^\circ$). To reduce the laser threshold we could leave just an unsilvered region arround $x=L/2$ to permit the escape of light only for $\chi_c\approx 33^\circ$. In this situation a refraction index of $n=1.85$ leads to a reflectivity of $\sim 70$ percent. The emission under these conditions is shown in Fig. 4(b).

The I-type QBS of Fig. 2(b), living in the large island centered at $x=L/2$ and with a small $\sin(\chi)$ in both PMs of Fig. 1 (as shown in \cite{ketz2}), can be directly associated with a curved mirror Fabry-Perot resonator since it bounces with almost normal incidence between both walls of the cavity. We will consider only the flat boundary of the cavity to be completely silvered, but the next arguments also work in the opposite case. To construct a kind of Fabry-Perot resonator with ``one semi-silvered mirror" we need a semiconductor material with $n\simeq 6$ in order to have a reflectivity of about $50$ percent ($R \sim 0.5$), see Fig. 4(c). But note that since $n$ is quite large, the beam spreads out due to Snell's law. This problem may be fixed by also silvering the deformed boundary of the cavity but leaving a hole smaller than the width (in real space) of the I-type resonance. For example, if the unsilvered hole is half of the width of the I-type function, the reflectivity will increase up to 65 percent. However, under this condition, already a value of $n=3.7$ produces a 50 percent reflectivity which corresponds to a Fabry-Perot resonator with a semi-silvered mirror, see Fig. 4(d).

In Ref. \cite{arc1,arc2} it was shown that the ray dynamics prediction of lasing emission is in rather good agreement with experimental realizations, so we expect that the predictions of Fig. 4 are also valid.\\

It is clear that by an appropiate choice of the geometrical parameters of the cavity, one can tailor design the emission pattern. For example, V-type QBS may be produced with $(d,a,L) = (1.0,0.51,5.55)$, as shown in Fig. 5(c). Figs. 5(a) and 5(b) are plots of the PMs for the deformed and flat boundaries of the cavity, respectively. In Fig. 5(c) the wave function of mode number 18 for a resonant energy (supporting 20 propagating modes) is shown. This is a V-type QBS living inside the stability islands of the PMs of Figs. 5(a-b). As before, if the flat surface of the cavity is silvered, the normal incidence of the V-type QBS on the deformed boundary will produce a kind of Fabry-Perot resonator with a ``semi-silvered mirror" when using a material with $n=6$ ($R \sim 0.5$), shown in Fig. 5(d). Finally, with a deformed silvered boundary, the reflectivity will be about 70 percent for an index of refraction $n=2.5$, see Fig. 5(e).

\begin{figure}[htb]
\begin{center}
\epsfig{file=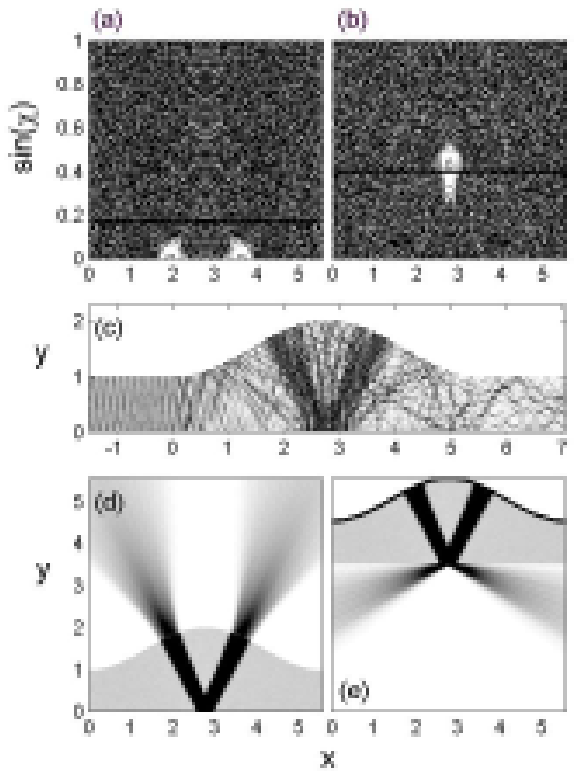,width=3.3in,height=3.7in}
\label{weaal0}
\end{center}
\end{figure}
\vspace{-0.28in}
{\small FIG. 5 Poincar\`{e} Maps for the closed cavity with surface of section (a) at the deformed wall, and (b) at the flat wall. $(d,a,L) = (1.0,0.51,5.55)$. Critical lines for refractive scape are shown. (c) V-type QBS $\mid \Psi(x,y) \mid^2$ corresponding to left-incomming mode 18, using $E=4095.1021$. Ray-optics prediction $\rho(x,y)$ of lasing emission for (d) $n=6$ and (e) $n=2.5$. The thick black line in (e) means a silvered surface.}\\

The quality factor $Q$ gives a measure of the response of the cavity to excitation. Following the procedure of Ref. \cite{po}, we obtain for our cavity the expresion $Q = 2 \pi \sqrt{R} n l / (1-R) \lambda$, where $l$ is the length of the trapped trajectory. The reflectivity $R$ depends on the index of refraction $n$ and, as in the case of Fig. 4(d), on the size of the unsilvered hole. For I-, V- and M-type QBS, $l$ is approximately $d+2a$, $2.2(d+1.57a)$, and $4.4(d+1.45a)$, respectively. Hence $Q$ is approximately four times larger for M-type that for I-type QBS, for a given $\lambda \ll l$.\\

In conclusion, we have proposed the experimental realization of microlasers as open semiconductor resonators with controlable laser emission. Using ray as well as wave dynamics we explained the physical mechanism by which lasing can occur; it takes advantage of mixed chaotic dynamics together with quantum tunneling into classically forbidden phase space regions. Finally, it is worth to point out that the values of the index of refraction used here are typical of conventional semiconductor materials and alloys \cite{crc}.\\

{\bf Acknowledgements:} This work was partialy supported by CONACyT, Mexico.

\end{multicols}


\begin{references}
\bibitem{note2} Although the waveguide is a scattering system, it is convenient, as it will be soon evident, to construct the PM assuming the cavity is closed at $x=0$, $L$.
\bibitem{note1} The standard set of variables that unambiguously specifies the PM for billiards are the Birkhoff variables, defined by the position and the angle of incidence with respect to the tangent at the collision point. However, following the conventional use (see Ref. \cite{arc1,arc2}), we plot, instead, the angle with respect to the normal. By doing so, the plots do not distinguish between both directions ($P_x<0$ and $P_x>0$). This is why Fig. 1(a), which we continue to call PM, only shows two islands representing the period-four resonance.
\end{references}
\end{document}